\documentclass[letter]{aa} 

\usepackage{graphicx}
\usepackage{subfig}
\usepackage{txfonts}
\usepackage[flushleft]{threeparttable}
\usepackage{mathtools}
\usepackage{multirow}
\usepackage{colortbl}
\usepackage{comment}
\usepackage{stfloats}
\usepackage{float}
\usepackage{afterpage}
\usepackage{hyperref}  
\hypersetup{colorlinks=true,linkcolor=[rgb]{0.1,0.4,1.},citecolor=[rgb]{0.1,0.4,1.},filecolor=[rgb]{0.7,0.2,0.2},urlcolor=[rgb]{0.1,0.4,1.}}

\usepackage{color}
\definecolor{blue}{rgb}{0., 0., 1}
\definecolor{lightblue}{rgb}{0.1,0.4,1.}

\usepackage{soul}
\usepackage{multicol}

\begin{document} 

\title{Impact of merger histories on the timing argument estimate of the Local Group mass} 

\titlerunning{Timing argument with merger histories}
\authorrunning{I. Akib et al.}

\author{Istiak Akib\inst{\ref{ins1}} \fnmsep\thanks{E-mail: \href{mailto:istiak-hossain.akib@obspm.fr}{istiak-hossain.akib@obspm.fr}}
\and François Hammer\inst{\ref{ins1}}
\and  Yanbin Yang\inst{\ref{ins1}}}

\institute{
LIRA, Observatoire de Paris, Universite PSL, CNRS, Place Jules Janssen 92195, Meudon, France \label{ins1}}

\date{Received 15 January 2025 / Accepted 14 February 2025}

\abstract{The timing argument (TA) aims to find the total mass of the Local Group (LG) from the relative motions of the Milky Way (MW) and Andromeda galaxy (M31). However, the classical TA always overestimates the LG mass, presumably because it does not account for the hierarchical scenario and other interactions, such as that with the Large Magellanic Cloud (LMC). We focused on the impact of M31's recent major merger by using three merger models to find the peculiar motion of M31 within the simple two-body, point-mass scenario of the TA. We find that the merger correction affects the TA mass by plus or minus 10-15\% depending on the M31 tangential motion, which has very large uncertainties. If we consider an M31 merger configuration that reduces the TA mass by 10-15\%, to which we add the impact due to the LMC infall into the MW as reported in the literature, the TA mass is consistent with the LG mass from Hubble-Lemaître flow. Galaxies are believed to have experienced about 16 major mergers each since z=11.5. Assuming all these mergers had a similar impact on the TA mass as the most recent M31 merger, the ratio of LG mass to TA mass would be $0.85^{+0.65}_{-0.37}$, and such a TA mass is consistent with all the LG mass estimates. Our result also agrees with findings that used LG analogues in cosmological simulations. We find that the TA mass estimate is limited by the hierarchical scenario, since it is not possible to track the progenitors of both the MW and M31 through so many mergers. We conclude that the MW--M31 dynamical system is far too complex to be modelled as a simple two-body point-mass system.}


\maketitle

\section{Introduction}
The Local Group (LG) is dominated by two large spiral galaxies, the Milky Way (MW) and the Andromeda galaxy (M31). One approach to finding their total mass is the so-called timing argument (TA), which relies on the fact that M31 is observed to be approaching us in the current epoch, while all other large galaxies have been moving away from us since the Big Bang. The two galaxies must have been moving away from each other during the Big Bang, and their gravity has since been bringing them towards each other. So, solving this simple two-body system using the current observed distance, velocity, and time would give us the total mass of the two galaxies. However, this classic calculation \citep{Kahn1959, lynden1981, vanderMarel2012} consistently predicts a much higher mass ($\sim 4-6\times10^{12} M_{\odot}$) than the LG mass ($\sim 2-3 \times 10^{12} M_{\odot}$) found from the sum of the highest mass estimates of the two galaxies from their internal or satellite dynamics \citep{Eilers2019, Wang2021, Chemin2009, Corbelli2010, Zhang2024, Watkins2010} or the dynamics of the LG as a whole \citep{Penarrubia2014, Diaz2014}. Previous attempts to reconcile the TA mass with the LG mass had to deviate from the two-body point-mass scenario by incorporating the impact of the Large Magellanic Cloud (LMC) on the MW \citep{Penarrubia2015, Benisty2024} or by using cosmological simulations to take the growth of the extended halos of the two galaxies over time into account \citep{Sawala2023, Benisty2024}.
 
However, the two-body point-mass scenario is affected by major merger events experienced by these galaxies in the past. In particular, we focus on the recent major merger experienced by M31, which is needed to reproduce the observational properties of M31 and its outskirts \citep{Hammer2018,Hammer2025, DSouza2018, Bhattacharya2023, Tsakonas2024}. Simulations by \cite{Hammer2018, Hammer2025} indicate that this merger started 7-10 Gyr ago and ended 2-3 Gyr ago. The observed position and velocity of M31 therefore have an additional component due to this merger, while the TA attributes the observation completely to the mutual gravity of the two galaxies.

In this Letter we show how this merger impacts the TA mass estimation. In Sect. \ref{sec: TA with merger} we recalculate the TA mass by correcting the global motion of M31 due to the merger using three M31 merger models. Section \ref{sec: discussion} shows that adding the merger correction to the reported LMC corrections can make the TA mass compatible with the LG mass, and we propose a correction related to the entire merger history of the two galaxies.

\section{Timing argument with the M31 merger correction}
\label{sec: TA with merger}
\begin{figure*}[ht]
\centering
    \includegraphics[width=1.8\columnwidth]{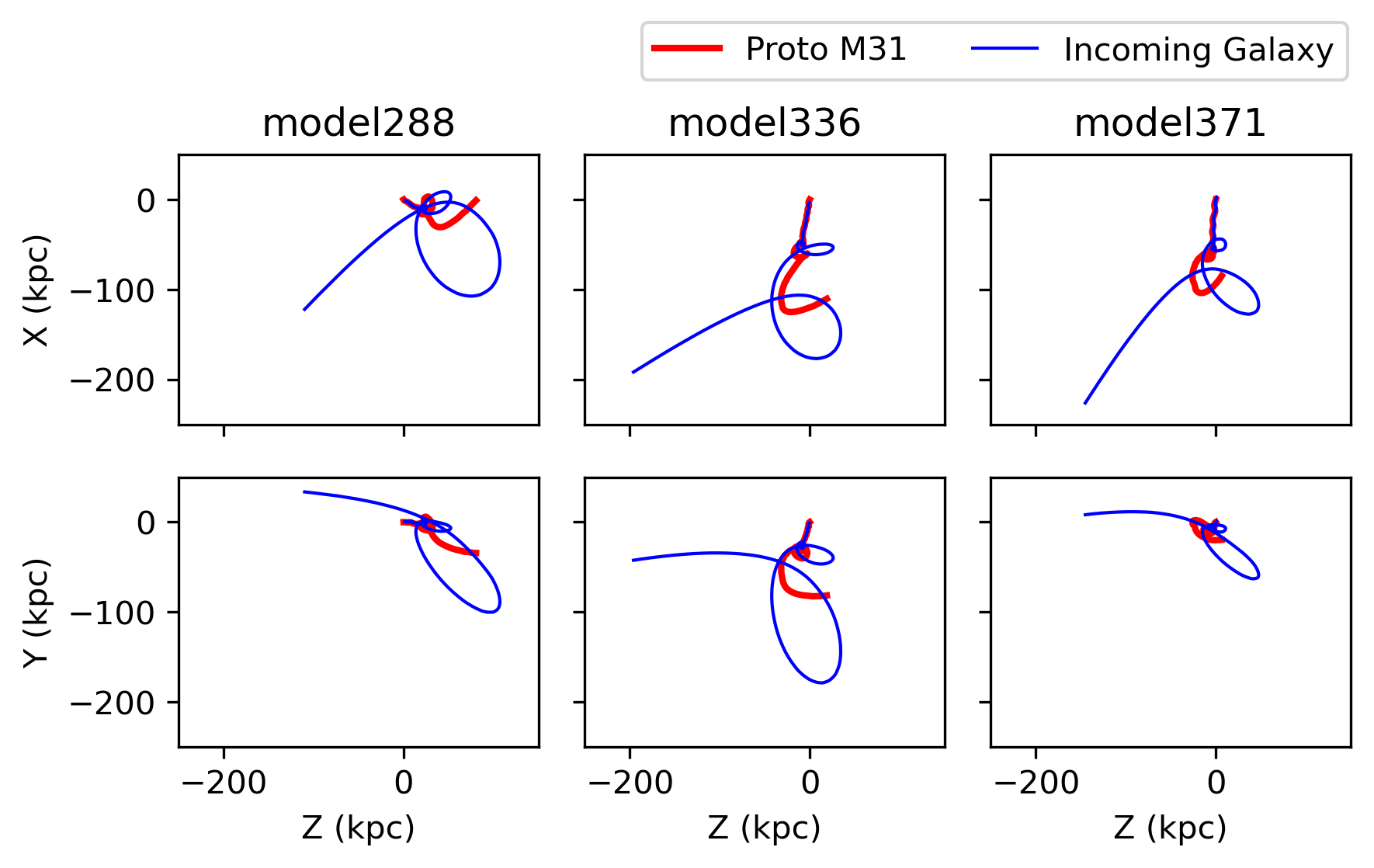}

 \caption[Merger Models]{Trajectories of the proto-M31 (red) and the incoming galaxy (blue) from three different M31 major merger models. The positions are shown in the Andromeda-centric reference frame; the current MW is on the positive side of the Z-axis.}
 \label{merger models}
\end{figure*}

\begin{figure}[hb!]
\centering
    \includegraphics[width=\columnwidth]{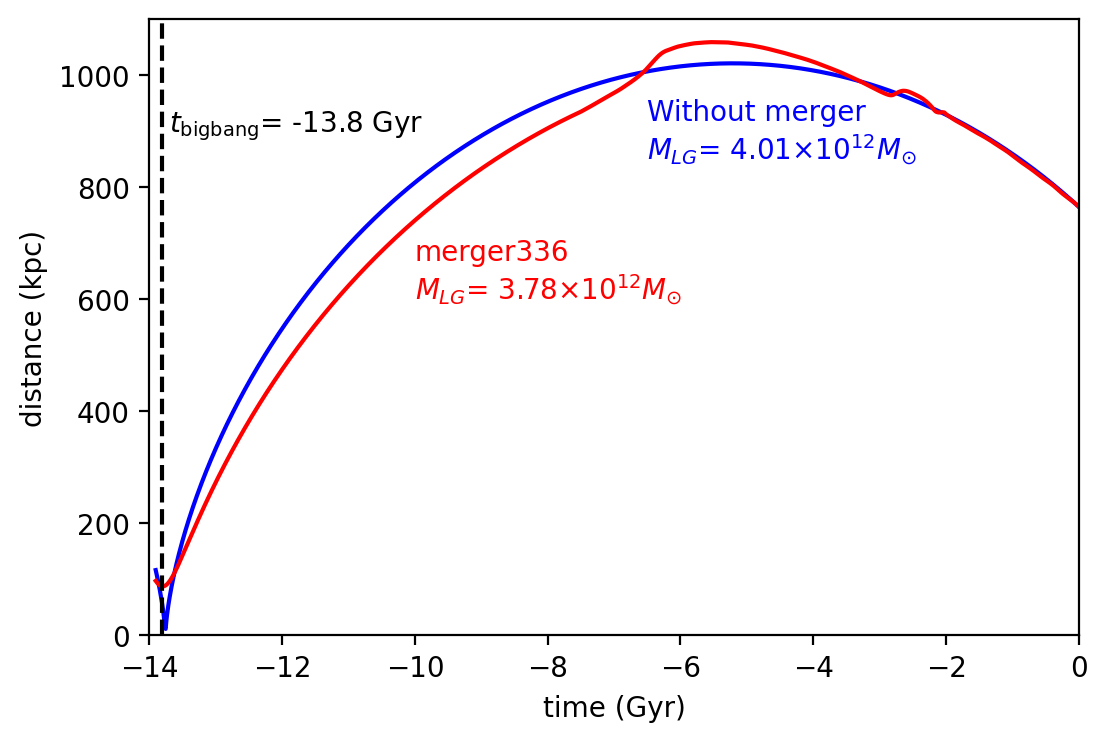}

 \caption[Merger correction]{Distance vs time plot for the TA mass calculation for the radial motion of M31 in the classical case (blue) and correcting for M31 merger model 336 (red).}
 \label{merger corr}
\end{figure}

We used M31 merger models 288 and 336 from \cite{Hammer2018} and model 371 from \cite{Hammer2025}. All three models have similar initial conditions and reproduce the observed properties of M31. In these merger models, a relatively light secondary galaxy, referred to as the incoming galaxy in this Letter, merges into a primary galaxy that is about four times more massive, which we refer to as the proto-M31. Model 371 is almost a copy of model 288, but the dark matter mass of both progenitors is reduced by factor of 1.6 in order to accurately fit the observed M31 rotation curve  \citep{Hammer2025}. Due to the lower masses, the orbit has been scaled down to a smaller pericentre (25 kpc) than that of model 288 (31 kpc) to match the star formation history of M31 \citep{Hammer2018}. The incoming galaxy of model 336 has a slightly different spin (by 15 degrees) than that of model 288 and a slightly larger orbital pericentre at 33 kpc. Figure \ref{merger models} shows the trajectories of the proto-M31 and the incoming galaxy. To calculate the TA mass, we numerically solved the two-body point-mass system under Newtonian gravity using initial conditions based on the present-day 3D positions and velocities of M31. We iteratively found a mass for the system that would create the minimum distance between the two galaxies at the Big Bang. To correct for the mergers, we took the 3D position and 3D velocity of the proto-M31 (in red) during this merger and added it to the global motion of the MW--M31 system over time. Figure \ref{merger corr} shows one such scenario by comparing the TA mass calculation for a radial approach (in blue) and when a correction for merger model 336 is adopted (in red). Since the merger started $\sim$ 6.5 Gyr ago and ended $\sim$ 2.5 Gyr ago in this model, we see small deviations in the distance between the two galaxies versus the time plot compared to the smooth line for the two-body system. In this particular case, the merger correction requires a lower LG mass than the simple TA calculation.

The initial conditions for the TA calculation are the position and velocity of M31. The 3D velocity of M31 comes from its observed radial velocity, proper motion (PM) with respect to the Sun, and the solar motion around the MW centre. The PM of M31 is very poorly constrained, which directly corresponds to a high uncertainty on the tangential motion of M31. For example, \cite{vanderMarel2019}, \cite{Salomon2021}, and \cite{Rusterucci2024} calculated the M31 PM from \textit{Gaia} DR2, EDR3, and DR3, respectively, with large uncertainties. \textit{Gaia} EDR3 systematics of 25 \textmu as yr$^{-1}$ would correspond to a 1$\sigma$ uncertainty of about 90 km s$^{-1}$ at the M31 distance \citep{Lindegren2021}. On the other hand, the radial velocity of M31 from the MW centre is about 110 km s$^{-1}$, and the tangential motion is consistent with 0. Another available M31 PM value that is widely used in the literature comes from \textit{Hubble} Space Telescope (HST) measurements \citep{vanderMarel2012}, which have a much smaller reported uncertainty than the \textit{Gaia} measurements. However, this measurement used three small fields of view (compared to the size of M31 in the sky), with only one field on the M31 disc \cite[see their Fig. 1]{Brown2006}. Furthermore, \cite{Brown2006} used an old model of M31 based on a single minor merger to correct for these three fields. These two issues would introduce significant uncertainty and possible inconsistency with our major merger model of M31, which has not been accounted for. So, in our analysis, we considered the M31 PM to be a free parameter and calculated the TA mass for a uniform grid of the M31 PM, which includes the observational estimates. Table \ref{Table parameters} lists the other observational parameters that we used in our calculation.

\begin{table}[h]
\centering
\begin{threeparttable}
\caption{\label{Table parameters} Observational values for the TA mass calculation.}
\centering
\begin{tabular}{|c|c|}
 \hline
 parameter& value\\
 \hline
 M31 distance (kpc)    & $761\pm11^a$\\
 M31 radial velocity (km $\mathrm{s^{-1}}$)  & $301\pm1^b$ \\
 Sun to the MW disc (kpc) & $8.275\pm0.042^c$ \\
 Sun vertical shift (kpc) & $0.0208\pm 0.0003^d$\\
 \multirow{2}{*}{Solar motion (km $\mathrm{s^{-1}}$)}
    &$[11.1\pm1.2,255.2\pm5.1,$ \\
    &$7.2\pm1.1]^e$  \\
 Age of the Universe (Gyr)  & $-13.801 \pm 0.02^f$\\
 \hline
\end{tabular}
\begin{tablenotes}
\item \textbf{References}: (a) \cite{Li2021}, (b) \cite{Bergh2000}, (c) \cite{Gravity2021}, (d) \cite{Bennett2018}, (e) \cite{reid2014}, (f) \cite{Planck2020}.
\end{tablenotes}
\end{threeparttable}
\end{table}

\begin{figure}[h!]
\centering
    \includegraphics[width=\columnwidth]{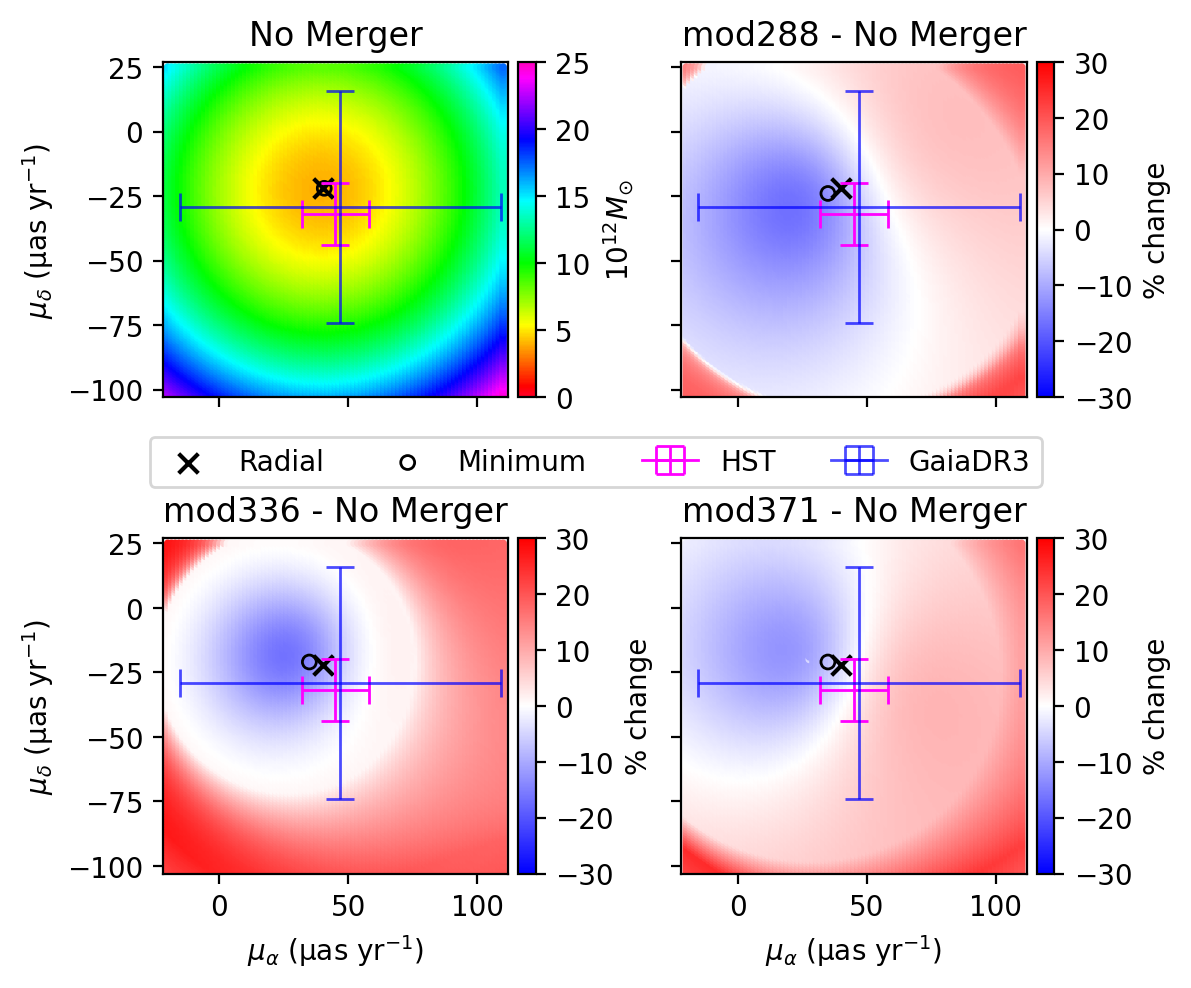}

 \caption[PM grid]{TA mass for different M31 PM values. The top-left panel shows the TA mass for the classical two-body scenario. The rest of the panels show the change from the classical case when position and velocity are corrected over time using M31 merger modelling. The cross and open circle indicate the M31 PM for the radial approach and minimum TA mass, respectively. The PMs of M31 from HST \citep{vanderMarel2012} and \textit{Gaia} DR3 \citep{Rusterucci2024} are also shown.}
 \label{pm Grid}
\end{figure}

Figure \ref{pm Grid} shows the impact of the merger and of the PM of M31 on the TA mass. In this calculation, we used only the mean values for the observational parameters. As expected, for the classic two-body TA mass calculation, the minimum mass (3.97$\times 10^{12} M_{\odot}$) is found for a completely radial motion between the two galaxies (top-left panel of Fig. \ref{pm Grid}). For a higher tangential motion, the TA mass is higher and is consistent with mass values as high as $\sim$ 15$\times 10^{12} M_{\odot}$ at 1$\sigma$ within the uncertainties from the \textit{Gaia} DR3 PM. Whether correcting for the M31 merger models would increase or decrease the TA mass depends on the M31 PM. The percentage change for the merger correction compared to the classic case ranges from a $\sim$20\% decrease to a $\sim$30\% increase within the calculation range across the three merger models. Within the uncertainties from \textit{Gaia} DR3 PMs, about half of the PM values result in a higher TA mass and the rest in a lower TA mass across the three merger models. The minimum TA mass values when correcting for models 288, 366, and 371 are 3.51$\times 10^{12} M_{\odot}$, 3.46$\times 10^{12} M_{\odot}$, and 3.72$\times 10^{12} M_{\odot}$, respectively, and these values are found close to the radial motion value (shown with open circles in Fig. \ref{pm Grid}). These mass values are significantly lower than the lowest mass value of the classical case (for the radial approach) but still higher than the expected LG mass from the perturbation of the Hubble-Lemaître flow, such as $ 2.3 \pm 0.7 \times 10^{12} M_{\odot}$ from \cite{Penarrubia2014}.

\section{Discussion}
\label{sec: discussion}
\begin{figure*}[tbp]
\centering
    \includegraphics[width=1.8\columnwidth]{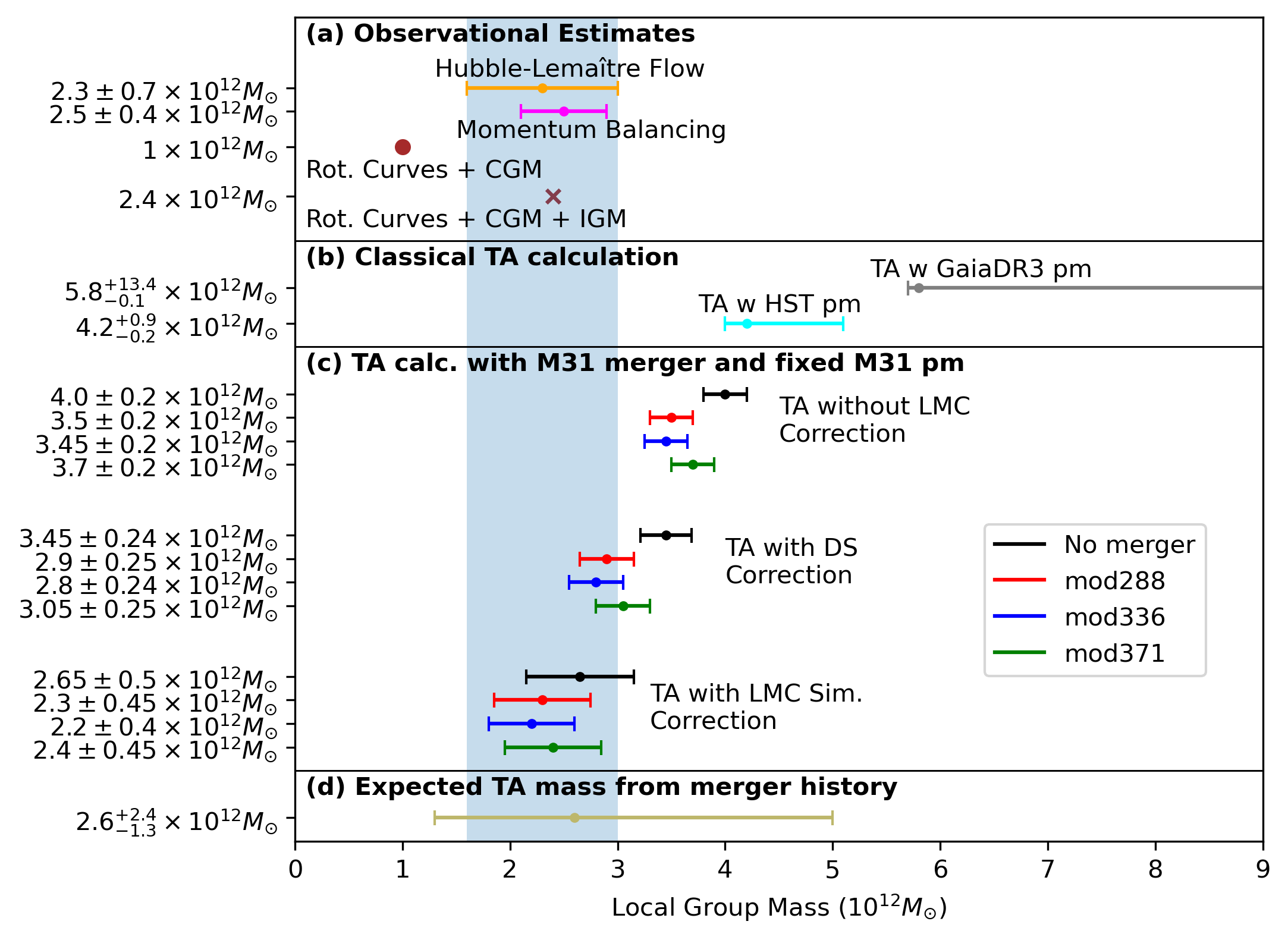}

 \caption[TA values]{Different LG mass estimates. Panel a: Observational estimates of the LG mass from the LG Hubble-Lemaître flow \citep[blue shaded region;][]{Penarrubia2014} and momentum balancing \citep{Diaz2014}. The sum of the MW \citep{Jiao2023} and M31 \citep{Hammer2025} mass from their rotation curves with the CGM \citep{Wang2019} and IGM \citep{McConnachie2007} mass is also shown. Panel b: TA mass with the M31 PMs from \textit{Gaia }DR3 \citep{Rusterucci2024} and HST \citep{vanderMarel2012}. Panel c: TA calculation with the M31 merger correction and the M31 PM fixed for the minimum mass. Three sets of TA calculations are shown: without LMC corrections, with disc-sloshing corrections \citep{Chamberlain2022}, and with corrections using LMC infall simulations \citep{Benisty2022}. Panel d: Expected TA mass assuming 16 major mergers for each galaxy.}
 \label{TA vals}
\end{figure*}

We have shown that correcting for the recent M31 major merger can significantly increase or decrease the TA mass estimate depending on the tangential motion of M31. The M31 tangential motion has very large uncertainties, and the PM values for the minimum mass are very close to the purely radial approach and consistent with observations (see Fig. \ref{pm Grid}). We want to determine whether the TA mass could be consistent with the LG mass. Thus, we considered only the PM values that minimise the TA mass and calculated the uncertainty of the TA mass estimate based on the uncertainties of the rest of the observational parameters (Fig. \ref{TA vals}, panel c). Accounting for the M31 merger may decrease the TA mass, but it is still significantly higher than the LG mass (Fig. \ref{TA vals}, panel a). We added the merger correction to two different corrections that account for the reflex motion of the MW due to the LMC infall reported in the literature, from \cite{Chamberlain2022} and \cite{Benisty2022}. \cite{Chamberlain2022} corrected the MW velocity based on the disc-sloshing effect observed in the halo stars by \cite{Petersen2021}. This constraint comes directly from the observed asymmetry in the velocity of halo stars compared to the MW centre, indicating a net travel velocity for the MW disc towards an earlier position of the LMC, which they corrected for in their TA calculation. On the other hand, \cite{Benisty2022} corrected the MW position and velocity using LMC infall simulations based on the method proposed by \cite{CorreaMagnus2021}. They calculated the change in position and velocity of the MW in a range of LMC infall simulations and determined the M31 position and velocity in the MW frame as if there were no LMC. The TA mass is consistent with the LG mass if we correct for both the M31 merger and LMC infall reflex and also set the M31 PM accordingly (Fig. \ref{TA vals}, panel c).

The observational LG mass estimates we show in panel a of Fig. \ref{TA vals} were found by modelling the LG Hubble-Lemaître flow \citep{Penarrubia2014} or by assuming a specific value of the LG momentum \citep{Diaz2014}. These estimates of the LG mass take other galaxies at distances of 1-3 Mpc from the LG barycentre into consideration. Figure \ref{TA vals} also shows the sum of the latest mass estimate found by fitting the rotation curve of the two galaxies. \cite{Jiao2023} found the MW mass to be $2.06^{+0.09}_{-0.06} \times 10^{11} M_{\odot}$ within a virial radius of $\sim$ 121 kpc by fitting the \textit{Gaia} DR3 rotation curve; this value\ is consistent with the MW mass determined using its accretion history \citep{Hammer2024}. \cite{Hammer2025} found the M31 mass to be $4.5 \times 10^{11} M_{\odot}$ within 137 kpc by fitting the observed HI rotation curve. There is a significant amount of gas mass outside the virial radii of these two galaxies in the forms of circumgalactic medium (CGM) and intergalactic medium (IGM) that affects the dynamics of the MW--M31 system. We took the CGM mass within 300 kpc of the MW from the modelling of \cite{Wang2019}: $1.2 \times 10^{11} M_{\odot}$. For the M31 CGM, we assumed its mass to be twice that of the MW CGM. Combining these values gives an estimate of $\sim 10^{12} M_{\odot}$. A constant IGM density of $4\times10^{-6}$ atoms/cm$^3$ \citep{McConnachie2007} within a sphere of radius = 1.5 Mpc would correspond to an IGM mass of $1.4\times10^{12} M_{\odot}$. The sum of the mass in the CGM and IGM and the masses from the MW and M31 rotation curve ($2.4\times10^{12} M_{\odot}$) is consistent with the LG mass estimates based on the Hubble-Lemaître flow or on the momentum balancing, as shown in panel a of Fig. \ref{TA vals}.

The MW had its last major merger 9-10 Gyr ago, as indicated by the observation of the Gaia Sausage Enceladus \citep{Belokurov2018, Haywood2018, Helmi2018}. Both the MW and M31 likely experienced many earlier mergers since the Big Bang, but the observational signatures of those earlier mergers have been washed out by the most recent one. It is thus not possible to track the full merger history of these two galaxies from observations and to use it to correct the TA mass calculation -- especially at the very early epoch just after the Big Bang. \cite{Duan2024} provide galaxy merger rates up to z=11.5 using observations from the \textit{James Webb} Space Telescope. According to their findings, between z=11.5 and z=6 (0.4 Gyr to 0.95 Gyr after the Big Bang), the merger rate was almost constant, about three major mergers on average per galaxy. From z=6 to z=3 (2.2 Gyr after big bang), there should be about 12 mergers. From z=3 to the current epoch, about one major merger is estimated -- which is what we observe for both the MW and M31. The merger mass ratio ($\leq$ 4:1) considered by \cite{Duan2024} includes that expected for the MW \citep{Naidu2021} and for M31, so we can assume that each merger has a similar impact on the TA mass estimates than what we found from the last M31 merger. Let us assume that the MW and M31 experienced 16 major mergers each since z=11.5 and that the change in the TA mass due to these mergers follows a normal distribution centred at 0 with a standard deviation of 10\% -- which is consistent with our findings for the latest M31 merger.  The ratio of the LG mass to the TA mass for such a scenario is $0.85^{+0.65}_{-0.37}$ (median with the 16th and 84th percentiles). Adopting the LG mass value from \cite{Penarrubia2014}, the expected TA mass would be $2.6^{+2.4}_{-1.3} \times 10^{12} M_{\odot}$. Hence, all TA mass estimates would be consistent within 1$\sigma$ (see Fig. \ref{TA vals}, panel d.)

How do the above corrections relate to the corrections from the cosmological simulations? \cite{Hartl2022} find that in the IllustrisTNG N-body and hydrodynamical simulation there is a bias towards the TA mass being overestimated compared to the virial mass of the MW--M31 analogues. \cite{Benisty2024} used this simulation to correct the TA mass and reach a level consistent with LG mass estimates. They find the ratio of the LG mass to the TA mass to be 0.82$\pm$0.2 for 600 LG analogues. This is consistent with our value of $0.85^{+0.65}_{-0.37}$ for 16 major mergers for each galaxy. Cosmological simulations start at the Big Bang and take the full merger history of each of the two galaxies  into account over the age of the Universe. These mergers can generate a large variety of directions between the tangential motion and the merger. However, we have shown that the TA mass is very sensitive to the tangential velocity. For example, \cite{Benisty2022} used different selection criteria for the radial and tangential velocity than \cite{Benisty2024} on the same cosmological simulation and find the ratio of the LG mass to the TA mass to be 0.63$\pm$0.2 for 160 LG pairs. Furthermore, the MW--M31 pair is unique in terms of their merger history: the MW's last merger is ancient, while M31's is recent, and the MW was within the orbital plane of M31's last merger \citep{Hammer2018}. Since there are further constraints based on the currently observed initial conditions of this system, reproducing a close enough MW--M31 analogue in cosmological simulations is essentially impossible. The results from the cosmological simulations therefore only provide a general view of the TA mass for pairs of galaxies in the Universe and may not be well suited to follow the specific case of the MW--M31 system.

In this Letter we have shown that correcting the M31 motion due to its recent major merger and the MW's reflex motion due to LMC infall can make the TA mass compatible with the LG mass. The most significant uncertainty in this argument comes from the large uncertainty associated with the M31 PM, which might be improved with subsequent \textit{Gaia} data releases. We can only correct for the most recent M31 major merger since its full hydrodynamic modelling is available, though both galaxies have likely experienced 16 collisions since the Big Bang. Assuming all these mergers had similar effects, we find that the classical TA mass estimate is consistent with the LG mass under such a scenario. Our finding is consistent with the results from cosmological simulations. The complex merger histories of both galaxies since the Big Bang and the recent LMC infall show that the dynamics of the MW--M31 system are far too complex to be modelled with a simple two-body point-mass system, as was previously proposed by the TA.

\begin{acknowledgements}
We would like to thank the referee for their insightful comments and suggestions, which greatly strengthened the overall manuscript. We thank David Benisty and Eugene Vasiliev for their very helpful discussion. Istiak Akib would like to thank the Graduate Program in Astrophysics of the Paris Sciences et Lettres (PSL) University for funding this research. 
\end{acknowledgements}

\bibliographystyle{aa}
\bibliography{ref}

\end{document}